\documentclass[prl,twocolumn,showpacs]{revtex4}
\newcommand{\be}{\begin{equation}}
\newcommand{\ee}{\end{equation}}
\newcommand{\bea}{\begin{eqnarray}}
\newcommand{\eea}{\end{eqnarray}}
\usepackage{graphicx}
\usepackage{epsfig}
\usepackage{bm}

\def\ev{\mbox{eV}}

\def\be{\beta}

\def\frac#1#2{{\textstyle{{#1}\over {#2}}}}

\def\lsim{\mathrel{\rlap{\lower4pt\hbox{\hskip1pt$\sim$}}
    \raise1pt\hbox{$<$}}}
\def\gsim{\mathrel{\rlap{\lower4pt\hbox{\hskip1pt$\sim$}}
    \raise1pt\hbox{$>$}}}
\def\sqr#1#2{{\vcenter{\vbox{\hrule height.#2pt
         \hbox{\vrule width.#2pt height#1pt \kern#1pt
         \vrule width.#2pt}
         \hrule height.#2pt}}}}

\begin{document}
\title{The Higgs portal and an unified model for dark energy and dark matter}

\author{O. Bertolami}
\affiliation{Instituto Superior T\'ecnico, Departamento de F\'{\i}sica \\
Av. Rovisco Pais 1, 1049-001 Lisboa, Portugal}

\thanks{Also at Centro de F\'{\i}sica dos Plasmas, IST}

\email{orfeu@cosmos.ist.utl.pt}

\author{R. Rosenfeld}
\affiliation{Instituto de F\'\i sica Te\'orica - UNESP \\ Rua
Pamplona, 145, 01405-900, S\~{a}o Paulo, SP, Brazil}

\email{rosenfel@ift.unesp.br}

\begin{abstract}
We examine a scenario where the Higgs boson is
coupled to an additional singlet scalar field which we identify with a quintessence field.
We show that this results in an unified picture of dark matter and dark energy,
where dark energy is the zero-mode classical
field rolling the usual quintessence potential and the dark matter candidate
is the quantum excitation (particle) of the field, which is produced in the universe due to its coupling
to the Higgs boson.
\end{abstract}
\preprint{IFT-P.xx/2007}
\preprint{IST}
\today
\maketitle


{\it Introduction.} The Higgs boson is the missing link of the standard model (SM) of the
electroweak interactions. Its search will be a top priority at the Large Hadron Collider
(LHC) whose operation starts next year. In addition to unraveling the mechanism of
electroweak symmetry breaking and the ensued mass generation for fermions and electroweak
gauge bosons, it may also be a portal to new physics hitherto hidden in a standard model
singlet sector \cite{portal}.

Proposals for singlet extensions of the SM are not new in the
literature. The simplest extension is the addition of a real
singlet scalar field which couples only to the Higgs doublet. This
could be called the minimal non-minimal standard model (MNMSM)
\cite{mnmsm}. The contribution of this new field to radiative
corrections only arise at the two-loop level and were studied in
Refs. \cite{radiative}.

More recently, the salient consequences of this simplest extension, both phenomenological
and cosmological, have been analyzed. The phenomenological implications can arise in two
ways: mixing of the singlet with the Higgs boson and the possibility of invisible decay of
the Higgs boson into two singlet bosons \cite{portal,mnmsm,stealth,wells,gabe}. On general
terms, a coupling of the Higgs doublet $H$ and the scalar singlet field $\Phi$ of the form
$\Phi H^\dagger H$ will generate a mixing between the physical Higgs boson and the singlet
after spontaneous electroweak symmetry breaking. In addition, a quartic coupling like
$\Phi \Phi H^\dagger H$ results in the possibility of Higgs boson decay into a pair of
singlets and, in the case of a non-zero vacuum expectation value of the singlet, a mixing
can also be induced.

The singlet-Higgs mixing will in general weaken the standard model
Higgs coupling to ordinary fields, making its discovery more
challenging \cite{kingman}. The invisible decay of the Higgs boson can be
detected through the weak gauge boson fusion at the LHC \cite{invisible}.

On the cosmological side, it was realized that the scalar singlet is practically stable
and therefore can provide a dark matter candidate, sometimes dubbed as phion
\cite{stealth,gabe,zee1,mcdonald,bento1,zee2,bento2,burgess}. Bounds on the phion-Higgs coupling can be
obtained from the constraints on its relic abundance.

In this letter we aim to study the consequences of identifying the scalar singlet not only
with the dark matter, but also with the dark energy field, responsible for the recent
stage of accelerated expansion of the universe. Coupling of the dark energy field to
neutrinos has led to models of variable-mass neutrinos \cite{neutrinos}. Models with
interaction between dark energy and dark matter have also been extensively studied
\cite{vamps}. The interaction between dark components is also a crucial issue in the
unified model of dark energy and dark matter, the generalized Chaplygin gas model, which
can be realized through a complex scalar field \cite{Bento2002,Bilic2002} or via a real
scalar field \cite{Kamenshchik2001,Bertolami2004}. Some of these interacting models also
lead to mass-varying dark matter and their effect on the relic abundance was discussed in
Ref. \cite{abundance}. Therefore, it seems natural to analyze models where the dark energy
couples to the Higgs field, which arise in the simplest extensions of the SM discussed
above.

There are many models with ultra-light particles, such as
models containing moduli fields or pseudo-Nambu-Goldstone bosons.
In these models, the associated fields of these ultra-light particles
could play the role of the quintessence field.
These particles should not be coupled to normal matter in order to avoid
new long range interactions \cite{carroll}. However, it remains to be investigated whether
the Higgs could offer a portal to this dark sector.


{\it A preliminary study.} In order to get some insight for the possible constraints
arising from coupling the Higgs to a very light scalar, let us consider first a definite
model where the fields are already in their minimum energy configuration (contrary to the
usual quintessence models, where the fields are displaced from their minimum). As an
example we will adopt the minimal spontaneously broken hidden sector model of Schabinger
and Wells \cite{wells} involving the usual Higgs doublet $H$ and a further singlet complex
scalar field $\Phi$ with a potential
\begin{equation}
V(\Phi,H) = -m_H^2 |H|^2 - m_\Phi^2 |\Phi|^2  + \lambda |H|^4 +
\rho |\Phi|^4  + \eta |H|^2 |\Phi|^2.
\end{equation}
The fields develop non-vanishing vacuum expectation values, $\langle |H|^2  \rangle =
v^2/2$ and $\langle |\Phi|^2  \rangle = \xi^2/2 $ and the in unitary gauge, the physical
fields $h$ (the Higgs boson field) and $\phi$ mix through a non-diagonal mass matrix due
to a non-vanishing $\eta$. The mixing angle $\omega$ between the two scalars is given by
\begin{equation}
\tan \omega = { \eta v \xi \over (\rho \xi^2 - \lambda v^2) + \sqrt{ ( \rho \xi^2 -
\lambda v^2)^2 + \eta^2 v^2 \xi^2} }~.
\end{equation}

We will consider fifth force constraints to derive bounds on the mixing $\omega$,
following Ref. \cite{DZ}. The mixing of the SM Higgs field with a light scalar induces
long range Yukawa-like interactions. Hence one would expect a Yukawa interaction with
strength $g_e \sin \omega$ for electrons and $g_N \sin \omega$ for nucleons, where $g_e  =
2.9 \times 10^{-6}$ is the electron Yukawa coupling and $g_N = 2.1 \times 10^{-3}$ is the
nucleon Yukawa coupling \cite{cheng}.

We consider now a non-relativistic test body of inertial mass $M$ placed in the
gravitational field of the Earth (with mass $M_E$) at a distance $r$ from its center such
that it undergoes an acceleration given by:
\begin{equation}
a = a_{gr} + a_{\phi}~,
\end{equation}
where
\begin{equation}
a_{gr} = {M_E \over M_{Pl}^2 r^2}~,
\end{equation}
is the usual Newtonian acceleration, $M_{Pl}$ is the Planck mass and
\begin{eqnarray}
a_{\phi} &=& {\omega^2 \over M r^2}[g_N^2  N_N^E N_N^t + g_N g_e (N_N^E N_e^t + N_e^E
N_N^t)  \\ \nonumber &+& g_e^2 N_e^E N_e^t]
\end{eqnarray}
is the extra acceleration due to a new force arising from $\phi$ exchange, as long as $m_\Phi r \ll 1$;
$ N_{(N,e)}^{(E,t)}$ is the number of nucleons or electrons in the Earth and in the
test body.

With these definitions one can find that the difference in acceleration between 2 test
bodies of distinct compositions
\begin{equation}
\varepsilon = 2 {|a_1 - a_2| \over |a_1 + a_2|}
\end{equation}
can be written as
\begin{equation}
\varepsilon = {M_{Pl} \over \bar{m}} \omega^2 g_N g_e \Delta f_p
\end{equation}
where $\bar{m}$ is the average nucleon mass and $\Delta f_p$ is the difference in isotopic composition
of the 2 test bodies:
\begin{equation}
\Delta f_p = {N_p^{(1)} \over N_p^{(1)} + N_n^{(1)}} - {N_p^{(2)} \over N_p^{(2)} +
N_n^{(2)}}
\end{equation}
with $N_{(p,n)}$ being the number of protons or neutrons in the test body.

For typical isotopic differences of the order of $\Delta f_p = {\cal O} (10^{-1})$ and
using the experimental result $\varepsilon < {\cal O} (10^{-13})$ \cite{baessler} we get
an estimate for upper bound on the mixing angle
\begin{equation}
\omega < {\cal O} (10^{-20})~.
\end{equation}
Hence the mixing is severely constrained by 5th force experiments, even when it occurs
only with the Higgs sector.
However, we should point out that this is expected in the model discussed, since
in the limit $\xi \ll v$ the mixing angle is given by
\begin{equation}
\omega \simeq {\eta \over 2 \lambda} {\xi \over v} \simeq \eta {m_{\phi} \over m_H}
\end{equation}
where we are assuming quartic couplings of the order of one. In order for this limit to
apply, the Compton length of the light scalar should be of the order of the Earth radius,
implying in  a mass of the order of $10^{-14}$ eV, to be compared with a Higgs mass of the
order of $100$ GeV. Therefore, we conclude that there are no limits on the mixing constant
$\eta$, which implies that it is possible that the Higgs may have a large invisible width
into very light particles.

It is also possible that these light scalars are dark matter candidates. Since in
this case there is no symmetry preventing $\phi$ decay, we must check if in fact they
can survive until today. The light scalar decays through its mixing with the Higgs boson.
If $m_\phi < 2 m_e$, the dominant decay rate is into gluons via a top quark triangle
($g_{\phi \bar{t} t} = \omega g_{h \bar{t} t}$), which can be readily estimated:
\begin{equation}
\Gamma(\phi \rightarrow gg) \simeq \omega^2 \left({\alpha_s \over \pi}\right)^2 {G_F
m_\phi^3 \over 36 \sqrt{2} \pi}
\end{equation}
which implies that
\begin{equation}
\Gamma(\phi \rightarrow gg)^{-1} \simeq {1 \over \omega^2 (m_\phi/\mbox{\footnotesize{1
eV}})^3} 10^{-5} t_U
\end{equation}
where $t_U \simeq 10^{17} s$ is the age of the universe. Hence, in order for the light scalar to be around today
we must require
\begin{equation}
 \omega^2 \left({m_\phi \over 1 ~eV}\right)^3 < 10^{-5}
\end{equation}
which is easily satisfied in the case considered here \cite{CommentKingman}.

Hence one should consider the bounds discussed in Ref. \cite{bento2}, whose
reasoning we repeat here as it shows that the scalar $\phi$ unfortunately cannot be a dark matter candidate.
In order to do that we consider the cosmological evolution of the light scalar field. If
$\eta$ is sufficiently small, the singlet decouples early in the thermal history of the
Universe and are diluted by subsequent entropy production. In Ref.~\cite{bento1}, we have
considered out-of-equilibrium singlet production via inflaton decay in the context of
$N=1$ Supergravity inflationary models (see. e.g. \cite{bento2} and references therein).

On the other hand, for certain values of the coupling $\eta$, it is possible that $\Phi$
particles are in thermal equilibrium with ordinary matter. In order to determine whether
this is the case, we make the usual comparison between the thermalization rate
$\Gamma_{th}$ and the expansion rate of the Universe $H$.
In Ref.~\cite{bento1} it was found that if $\eta > 10^{-10}$ the singlets can be brought
into thermal equilibrium right after the electroweak phase transition, being as abundant as photons.

 Since we are interested in a stable ultra-light singlet field, it is
a major concern avoiding its overproduction if it decouples while relativistic. In fact,
in this case, there is  an analogue of Lee-Weinberg limit for neutrinos (see e.g.
\cite{Kolb}):

\begin{equation}
\Omega_\phi h^2 \simeq 0.08 \left({m_\phi \over 1 ~ \ev}\right) \quad,
\end{equation}
which shows that as $m_{\phi} \simeq 10^{-14}$ eV, the thermally produced ultra-light
scalar $\phi$ particles cannot be a dark matter candidate, as for it is required that
$\Omega_\phi h^2 \simeq 0.1$. Of course this is only a toy model since it requires a huge
fine-tuning to generate the different scalar mass scales and it should be embedded in a
more encompassing model, such as the Minimal Supersymmetric Standard Model with the
addition of one singlet chiral superfield \cite{BCFW}.

In what follows we shall consider the case where the Higgs is coupled to a quintessence
field and pay special attention to the fact that it is not settled in its minimum. It is
in this respect that a realistic dark energy candidate differs from the generic dark
matter candidate modeled by a scalar $\Phi$ considered above.


{\it Coupling the Higgs to quintessence.} Here one must carefully devise a potential
consistent with the classical equation of motion for the quintessence field $\Phi$,
avoiding the presence of unphysical tadpoles. We choose a generic potential of the form:
\begin{eqnarray}
V(\Phi,H) &=&  U(\Phi) + \lambda \left( |H|^2 - {v^2 \over 2} \right)^2 \\ \nonumber && +
\lambda_1 (\Phi - \Phi_{cl})^2 |H|^2 + \mu (\Phi - \Phi_{cl}) \left( |H|^2 -{v^2 \over 2}
\right)
\end{eqnarray}
where $U(\Phi)$ is the usual quintessential potential which can, for instance, be of the
form of an exponential or an inverse power law in the singlet field \cite{review}. The
second term is the SM Higgs potential and the last two terms give rise to
Higgs-quintessence coupling, where $\lambda_1$ is a dimensionless coupling constant and
$\mu$ a constant with dimensions of mass. The classical field $\Phi_{cl}$ is the solution
to the usual equation of motion:
\begin{equation}
\ddot{\Phi}_{cl} + 3 H \dot{\Phi}_{cl} + {d U \over d \Phi_{cl}} = 0.
\end{equation}

The time-dependent minimum of the potential is at
$\langle |H|^2 \rangle = \frac{v^2}{2}$ and $\langle \Phi \rangle = \Phi_{cl}$
where all energy density is due to $U(\Phi_{cl})$, exactly as standard quintessence models.
The Higgs doublet in unitary gauge can be written as
$H^\dagger = \left(0, \frac{1}{\sqrt{2}} (v + h) \right)$, where $h$ is the physical Higgs
boson and $v=246$ GeV is the vacuum expectation value.
The quantum fluctuations of the singlet field are described by
\begin{equation}
\varphi = \Phi - \Phi_{cl} ~,
\end{equation}
so that the accelerated
expansion of the universe is driven by the zero-mode classical
field $\Phi_{cl}(t)$, which
corresponds to the spatially homogeneous vacuum state and
typically is of the order of $M_{Pl}$.

 It is interesting to notice that this vacuum state does not arise from a mechanism of spontaneous
symmetry breaking but is instead a consequence of initial conditions for the slow rolling
evolution of the quintessence field, that is
\begin{equation}
{\dot{\Phi}_{cl} \over \Phi_{cl}} << H ~.
\end{equation}

In terms of the Higgs boson $h$ and the quintessence field $\varphi$ the potential becomes
\begin{eqnarray}
V(\varphi,h) &=& U(\Phi_{cl}) + \lambda v^2 h^2 + \lambda v h^3 + \frac{\lambda}{4} h^4
 \\ \nonumber &+&
\lambda_1 \varphi^2 h^2 + \frac{1}{2} \mu \varphi h^2 + 2 \lambda_1 v \varphi^2 h  + \mu v
\varphi h + \lambda_1 v^2 \varphi^2,
 \label{potential}
\end{eqnarray}
where we have taken $U(\Phi) \simeq U(\Phi_{cl}) $ since $\varphi \ll \Phi_{cl}$. The
$\mu$ term is responsible for the Higgs boson mixing with the quintessence particle. Hence
we have the interesting possibility of unifying dark energy and dark matter by imposing a
symmetry $\varphi \rightarrow -\varphi$, which requires $\mu = 0$ and renders the
quintessence particle stable. Furthermore, since the original quintessence field is not
coupled to ordinary matter and there is no mixing with the Higgs boson, there are no
bounds arising from 5th force constraints. Without mixing the Higgs boson mass is the
usual SM result $m_h^2= 2 \lambda v^2$ and the $\varphi$ excitation mass is $m_\varphi^2
=2 \lambda_1 v^2$.

In this unified picture, dark energy is the zero-mode classical field $\Phi_{cl}(t)$
rolling down the usual quintessence potential and the dark matter candidate is the quantum
excitation (particle) $\varphi$, which is produced in the universe due to its coupling to
the Higgs boson.

At this point one might get concerned about the effects of radiative corrections for the
quintessence potential $U(\Phi)$. Quantum fluctuations arising from the self-coupling of
the quintessence field have been shown to be harmless for potentials of the form of
inverse-power law, exponential and cosine-type in the quintessence field \cite{DJ}. The
one-loop induced renormalized effective potential arising from the interactions of the
quintessence field with other fields has been analyzed in Ref. \cite{Garny}. In there, the
coupling has a general mass dependence of the mass of the extra particles on the
quintessence field. In our case we see that in fact the Higgs mass is independent of
$\Phi_{cl}$ and therefore there are no contributions to the effective potential.

The analysis of the dark matter contribution arising from this singlet follows that of
Ref. \cite{bento2}. In particular, if the scalar decouples relativistically one needs to
have $m_\varphi \simeq 4$ eV in order to not over-close the universe. In this case, the
coupling to the Higgs boson, which is the same that gives mass to $\varphi$, must be tiny,
$\lambda_1 = {\cal O} (10^{-22})$. It is interesting that even such a small coupling can
lead to a successful model of dark energy and dark matter. Unfortunately, this tiny
coupling is most likely impossible to test at future accelerators, since it leaves the
Higgs sector of the SM practically unaffected.

On the other hand, there is another solution to the DM abundance that requires a large
value of coupling constant. Indeed, the solution of the Boltzmann equation studied in Ref.
\cite{bento2} admits a solution for which $m_\varphi \simeq 1$ GeV for $\lambda_1 = {\cal
O} (10^{-1})$, however this is clearly incompatible with the fact that $m_\varphi^2 =2
\lambda_1 v^2$.


{\it Conclusions.} In summary, we have examined the implications of a scenario where the
Higgs boson is coupled to a SM singlet field responsible for the accelerated expansion of
the universe. In this context a quite interesting possibility arises where the classical
zero-mode component of the singlet field corresponds to the dark energy particle while its
excitation plays the role of dark matter. In order to make this excitation consistent with
the cosmological density requirement $\Omega_\varphi h^2 \simeq 0.1$ implies the coupling
with the Higgs field is rather small $\lambda_1 = {\cal O} (10^{-22})$, which makes
unlikely that this scenario might be scrutinized in the forthcoming generation of
accelerators. Nevertheless, it is remarkable that a such a tiny opening in the Higgs
portal admits such a rich and intriguing scenario.


{\it Acknowledgments.} O.B. would like to thank the hospitality of the Department of
Astronomy and Astrophysics at the Fermi Institute of the University of Chicago where part
of this work has been written. O.B. also acknowledges the partial support of Funda\c c\~ao
para a Ci\^encia e a Tecnologia (Portugal) under the grant POCI/FIS/56093/2004. R.R. thanks
CNPq (grant 309158/2006-0) and FAPESP (grant 04/13668-0) for partial support. The authors
are indebted to Kingman Cheung, Vikram Duvvuri, Mathias Garny, Rocky Kolb and Christof Wetterich for many
important discussions.


\begin{thebibliography}{11}
\bibitem{kingman}
If the singlet is heavier than a pair of $b$ quarks, which is not the case we have in
mind, the Higgs decay would not be invisible.
See K.~Cheung, J.~ Song, Q.-S.~Yan,  Phys. Rev. Lett. {\bf 99}, 031801 (2007).
\end{thebibliography}
\end{document}